\documentclass[showpacs,preprint,amsmath,amssymb]{revtex4}
\usepackage{graphicx}
\usepackage{amsmath}
\usepackage{color}

\begin{document}
\title{Effects of free will and massive opinion in majority rule model}
\author{Zhi-Xi Wu}\thanks{wupiao2004@yahoo.com.cn}
\author{Guanrong Chen}\thanks{gchen@ee.cityu.edu.hk}
\affiliation{Department of Electronic
Engineering, City University of Hong Kong, Hong Kong SAR, People's
Republic of China}
\date{Received: date / Revised version: date}
\begin{abstract}
We study the effects of free will and massive opinion of
multi-agents in a majority rule model wherein the competition of
the two types of opinions is taken into account. To address this
issue, we consider two specific models (model I and model II)
involving different opinion-updating dynamics. During the
opinion-updating process, the agents either interact with their
neighbors under a majority rule with probability $1-q$, or make
their own decisions with free will (model I) or according to the
massive opinion (model II) with probability $q$. We investigate
the difference of the average numbers of the two opinions as a
function of $q$ in the steady state. We find that the location of
the order-disorder phase transition point may be shifted according
to the involved dynamics, giving rise to either smooth or harsh
conditions to achieve an ordered state. For the practical case
with a finite population size, we conclude that there always
exists a threshold for $q$ below which a full consensus phase
emerges. Our analytical estimations are in good agreement with
simulation results.
\end{abstract}
\maketitle

For a long time in history simple statistical models are used to
study complex biological, social, and geological phenomena
\cite{Stauffer2006book,Stauffer,Galam1998epjb}. Recently, a great
deal of efforts are devoted to mathematical modelling of
collective behaviors of individuals, particularly opinion
spreading and formation, by using for example two-state
interacting spin models
\cite{Galam1999pa,Sznajd2000ijmpc,Krapivsky2003prl,Li2006pre,
Castellano2007,Cao2008pre}. In this context, the common
formulation assumes that the agents are located at the nodes of a
network, and are endowed with two states, i.e., spin up and spin
down, which mimic human attitudes or decisions: like/dislike,
agree/disagree, accept/reject, etc. The interactions are assumed
to take place only between linked nodes, where the links represent
the relations between individuals such as friendship,
coauthorship, neighborhood, etc. Through interactions, the agents
can update their opinions by ways of convincement
\cite{Sznajd2000ijmpc}, imitation \cite{Castellano2007}, following
local majority \cite{Galam1999pa,Krapivsky2003prl,Li2006pre}, and
so on. Generally, the main concern of these models is whether
there arises an order-disorder phase transition of the system
dynamics \cite{Galam1999pa,Sznajd2000ijmpc} and, if yes, how the
time needed to attain consensus depends on the system size
\cite{Krapivsky2003prl}, the initial configuration
\cite{Sznajd2000ijmpc,Krapivsky2003prl,Cao2008pre}, or the
topology of the underlying network \cite{Li2006pre,Cao2008pre}.

Very recently, some extensions of the majority rule model (MRM)
\cite{Galam1999pa,Krapivsky2003prl} were proposed and investigated
within different scenarios
\cite{Lambiotte2007epl,Lambiotte2007pre,Guan2007pre}. In Ref.
\cite{Lambiotte2007epl}, Lambiotte studied a variant of the MRM on
heterogeneous networks, namely \emph{dichotomous networks}, which
are composed of two kinds of nodes characterized by their distinct
link degrees, $k_1$ and $k_2$. It was found that the degree
heterogeneity (characterized by the ratio $\gamma=k_1/k_2$)
affects the location of the order-disorder phase transition point
and that the system exhibits non-equipartition of average opinion
between the two kinds of nodes. The effects of community structure
on opinion spreading and formation in the framework of MRM was
also considered by Lambiotte and coworkers
\cite{Lambiotte2007pre}. Motivated by the fact that a social
system is inhomogeneous in many aspects of its inherent nature,
Guan \emph{et al.} introduced two types of agents in the MRM where
one type of agents have less ability to persuade the other
\cite{Guan2007pre}. It was shown that, as the inhomogeneous effect
is strengthened, the location of phase transition point is shifted
along the direction where the ordered state is more difficult to
be realized.


In the present paper, we continue the research in line of Ref.
\cite{Guan2007pre} to study the effects of free will and massive
opinion of multi-agents in the MRM. The motivation comes from the
following observations. In some cases, our opinions are strongly
influenced by our social surroundings (the opinions of our
friends, colleagues and neighbors), e.g., the fact that a majority
of our acquaintances are smokers will easily convince ourselves to
smoke; the fact that a large number of our friends have MySpace
will likely urge us to apply for one account. This phenomenon can
be modelled by implementing a local majority rule when the agents
update their opinions. In other cases, however, we make decisions
in a way less dependent on the others, such as the type of coffee
we like to drink, the fashion of clothes we like to wear, etc.,
where our taste, or ``free will", dominates. In some other cases,
not only the local majority opinion but also the global one would
affect much of our behavior. Therefore, it is important to
investigate how the free will and massive opinion of multi-agents
influence the processes of their opinion spreading and formation.

To proceed, we first introduce our model of opinion dynamics. The
population is composed of $N$ agents located on a fully connected
network. This simple setting allows us to look for a mean-field
analytical solution for the problem. Initially, two types of
opinions, $S_1$ and $S_2$, are assigned to the agents with an
equal probability. The densities of populations holding opinions
$S_1$ and $S_2$ are denoted by $\rho_1$ and $\rho_2$,
respectively. We investigate two types of models (model I and
model II) involving two different opinion-updating dynamics. At
each time step, one agent is randomly selected to update its
opinion state. Following Refs.
\cite{Lambiotte2007epl,Lambiotte2007pre,Guan2007pre}, two types of
processes may take place: $(\texttt{i})$ With probability $(1-q)$,
the selected agent interacts with two neighboring agents randomly
selected from its neighborhood, and the three agents adopt the
opinion of the local majority. The magnitude of $(1-q)$
characterizes the frequency of confrontation, or the strength of
aggressiveness \cite{Galam1998epjb}. This step is compulsory for
both models I and II. $(\texttt{ii})$ With probability $q$, the
selected agent picks an opinion by using one of the following two
different strategies.

In model I, we assume that where there is no
neighboring-interaction happening, an agent updates its opinion
state according to its own will. In particular, the selected agent
may change its current opinion to the opposite one with a free
will whose strength is weighed by $\alpha\in[0,1]$. Smaller values
of $\alpha$ implies stronger confidence for the agent to stick to
its current opinion. In model II, we assume that during updating,
the selected agent has a tendency to adopt the global massive
opinion with a probability proportional to its corresponding
population density in the system:
\begin{equation}
P(S_1)=\frac{\rho_1^\beta}{\rho_1^\beta+\rho_2^\beta}\nonumber,
\end{equation}
where $\beta$ characterizes the strength of the massive opinion.
Larger values of $\beta$ correspond to stronger tendency to become
the majority. Hereafter, we always denote the global massive
opinion by $S_1$. When there exist only neighboring interactions,
i.e., in the case of $q=0$, it is already known that the system
asymptotically reaches global consensus where all agents share the
same opinion \cite{Krapivsky2003prl}. In the following, we
investigate how the phase diagram of the system varies when the
free will and massive opinion of agents are taken into account in
the case of $q\neq0$. In this case, the quantity $q$ serves as the
control parameter in the dynamical process over the network.

For model I, it is easy to write down the following mean-field
rate equation for the system:
\begin{equation}
A_{t+1}=A_t+(1-q)W+q[\alpha(1-a)-\alpha a],
\end{equation}
where $A_t$ is the average number of agents with opinion $S_1$ at
time $t$, $a=A_t/N$ is the corresponding average proportion of
agents with this opinion, and $W$ is the total contribution to the
evolution of $A_t$ due to neighboring interactions. The term
proportional to $(1-q)$ accounts for local majorities and the last
term, for the flips with free will. The probability for two agents
with opinions $S_1$($S_2$) and one with $S_2$($S_1$) to be
selected is $3a^{2}(1-a)$ and $3a(1-a)^{2}$, respectively, so that
\begin{equation}
W=3a^{2}(1-a)-3a(1-a)^{2}=-3a(1-3a+2a^{2}).\label{interaction}
\end{equation}
So, the evolution equation for $A_t$ becomes
\begin{equation}
A_{t+1}=A_t+(1-q)[-3a(1-3a+2a^{2})]+q\alpha(1-2a)\label{rateeq}.
\end{equation}

\begin{figure}[b]
\begin{center}
{\includegraphics[width=8cm]{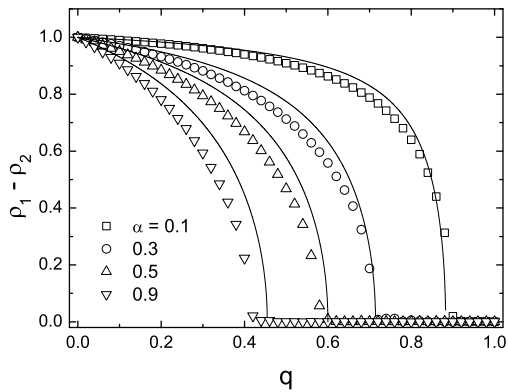}} \caption{The order
parameter $\rho_1-\rho_2$ versus the control parameter $q$ for
several values of $\alpha$. The solid lines are the analytical
solutions obtained by calculating Eq. (\ref{result1}). The symbols
are the simulation results obtained from averaging over twenty
independent experiments within a population of size
$N=5\times10^3$.} \label{fig1}
\end{center}
\end{figure}

Due to the existence of symmetry for the two opinions, it is easy
to see that $a=1/2$ is always a trivial stationary solution of the
above equation. In order to find nontrivial solutions of the
system, we use the method proposed in \cite{Lambiotte2007epl}.
Instead of considering directly the quantity $A_t$, we rewrite Eq.
(\ref{rateeq}) as follows and consider the quantities
$\vartriangle=A-N/2$ and $\delta=a-1/2$:
\begin{equation}
\vartriangle_{t+1}=\vartriangle_t+\frac{\delta}{2}
[3-(3+4\alpha)q-12(1-q)\delta^2].
\end{equation}
We can easily verify that the symmetric solution $a=1/2$ ceases to
be stable when $q<3/(3+4\alpha)$, and in that case the system
reaches the following asymmetric solutions:
\begin{equation}
\rho_{1,2}=a_{1,2}=\frac{1}{2}\pm\sqrt{\frac{3-(3+4\alpha)q}{12(1-q)}}.
\end{equation}
For convenience, we define the order parameter of the system by
the difference between the average density of the majority and
that of the minority, in the steady state, as follows:
\begin{equation}
\rho_1-\rho_2=\sqrt{\frac{3-(3+4\alpha)q}{3(1-q)}}\label{result1}.
\end{equation}
Thus, the system undergoes an order-disorder transition at the
critical point
\begin{equation}
q_c(\alpha)=\frac{3}{3+4\alpha}.
\end{equation}
Below this value, one type of opinion dominates the other, or an
ordered collective phenomenon emerges due to the interactions
between neighboring agents. According to Eq. (\ref{result1}), the
location of the order-disorder phase transition point depends
strongly on the value of $\alpha$, i.e., the strength of the free
will of the agents to change their current opinion states. The
stronger the inclination (to make a change) is, the more difficult
an ordered collective behavior emerges. Evidently, model I is
equivalent to the MRM considered in \cite{Lambiotte2007epl} with
$\alpha=0.5$, and in that case, $q_c=3/5$.

To test the above analysis, we perform computer simulations on
model I and compare the simulation results with the analytical
estimations in Fig. \ref{fig1}. Simulations were carried out for a
population of $N=5\times10^3$ agents located on the sites of a
fully connected network. We study the order parameter
$\rho_1-\rho_2$ as a function of the control parameter $q$.
Initially, the two opinions $S_1$ and $S_2$ are randomly
distributed among the agents with equal probability $1/2$. After
evolution, the system reaches a dynamic equilibrium state where
the densities of the opinions fluctuate stably. In our
simulations, one Monte Carlo step is accomplished after all agents
have updated their opinions. The simulation results were obtained
by averaging over the last $10^4$ Monte Carlo steps out of $10^5$.
From Fig. \ref{fig1}, we can see that the simulation results are
in very good agreement with the analytical solutions. The small
differences between the simulation results and the analytical
predictions are due to the finite-size effect.

\begin{figure}[t]
\begin{center}
{\includegraphics[width=8cm]{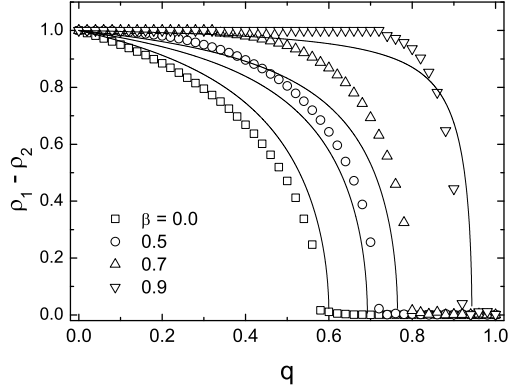}} \caption{The order
parameter $\rho_1-\rho_2$ versus the control parameter $q$ for
several values of $\beta$. The solid lines are the analytical
solutions obtained by calculating Eq. (\ref{result2}). The symbols
are the simulation results obtained from averaging over twenty
independent experiments within a population of size
$N=5\times10^3$.} \label{fig2}
\end{center}
\end{figure}

Now, we consider model II, for which the mean-field rate equation
for $A_t$ is given by
\begin{equation}
A_{t+1}=A_t+(1-q)W+q\frac{(1-a)a^\beta-a(1-a)^\beta}
{a^\beta+(1-a)^\beta}\label{rateeq2},
\end{equation}
where the formulation of $W$ is the same as that in model I. As
assumed above, the global majority opinion has a greater
attraction than the minority one, therefore we restrict our
attention to the region of $\beta\in[0,1)$ \cite{Note}. The last
term in Eq. (\ref{rateeq2}) accounts for the flips influenced by
the massive opinion, i.e., the global majority opinion of the
population. Since the average density of the majority opinion $a$
is less than unity (except for the full consensus case where
$a=1$), we can approximately solve Eq. (\ref{rateeq2}) by using of
the following truncated Taylor expansions:
\begin{eqnarray}
&(1-a)^\beta=1-\beta a+\mathcal{O}(a^2),\cr &
a^\beta=1-\beta(1-a)+\mathcal{O}(a^2).\label{taylor}
\end{eqnarray}
Substituting (\ref{interaction}) and (\ref{taylor}) into
(\ref{rateeq2}), with some algebraic calculations, we get
\begin{equation}
A_{t+1}=A_t+(1-q)[-3a(1-3a+2a^{2})]+q\gamma(1-2a),
\end{equation}
where $\gamma=(1-\beta)/(2-\beta)$.

\begin{figure}[b]
\begin{center}
{\includegraphics[width=8cm]{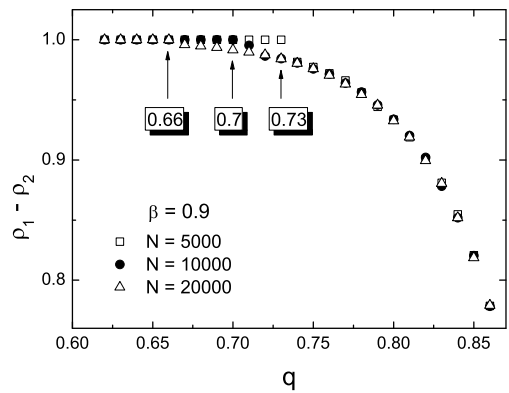}} \caption{The order
parameter $\rho_1-\rho_2$ versus the control parameter $q$ for
different system sizes $N$. The parameter $\beta=0.9$ is fixed.
The arrows point to the locations of phase transition obtained by
simulations.} \label{fig3}
\end{center}
\end{figure}

The remaining part is the same as what has been done for model I.
We hence omit the lengthy calculations and simply write out the
final result for the order parameter of model II:
\begin{eqnarray}
\rho_1-\rho_2&=&\sqrt{\frac{3-(3+4\gamma)q}{3(1-q)}} \cr &=&
\sqrt{\frac{3-[3+4(1-\beta)/(2-\beta)]q}{3(1-q)}}.\label{result2}
\end{eqnarray}
Thus, in model II, the system undergoes an order-disorder
transition at the critical point
\begin{equation}
q_c(\beta)=\frac{3(2-\beta)}{3(2-\beta)+4(1-\beta)} =
\frac{6-3\beta}{10-7\beta}.
\end{equation}
For $\beta=0$, our model reduces to the MRM studied in
\cite{Lambiotte2007epl}, so it is not surprising to recover the
known result of $q_c=3/5$. While for other positive values of
$\beta$ less than unity, we obtain a threshold of $q$ (at which
the order-disorder transition takes place) greater than $3/5$.
That is, if the information of global massive opinion is
accessible to the agents who have tendency to becoming the
majority, it is easier for the whole system to attain an ordered
state.

The simulation results of the order parameter $\rho_1-\rho_2$ as a
function of the control parameter $q$ for several values of
$\beta$ are shown in Fig. \ref{fig2}. The analytical solutions of
Eq. (\ref{result2}) are also plotted there for comparison. Once
again, they well match with each other. The differences between
the simulation and analytical results come from two aspects: one
is the finite-size effect, and the other is the approximation of
Eq. (\ref{taylor}). From Fig. \ref{fig2}, we also notice that for
any large enough value of $\beta$, there arises another phase
transition where a full consensus state emerges as the parameter
$q$ goes below some critical value $q_c^\prime$, i.e., one of the
two opinions dominates the whole system for any $q<q_c^\prime$.

We argue that the fact that the considered system is of finite
size contributes to the phase transition. In fact, reviewing Eq.
(\ref{rateeq2}), we find that $A_{t+1}$ will be going to $N$ iff
$q$ satisfies the following condition:
\begin{equation}
q<f(a)=\frac{3a(1-a)(2a-1)}{3a(1-a)(2a-1)+\frac{(1-a)a^\beta-a(1-a)^\beta}
{a^\beta+(1-a)^\beta}}.
\end{equation}
Note that $a$ is the average density of the majority,
$a\in[0.5,1]$. In this region of $a$, it is easy to verify that
$f(a)$ is a monotonically decreasing function for positive
$\beta$. Thus, as long as $q$ is smaller than the minimal value
$f_{min}$ of $f(a)$, the whole system will evolve to a full
consensus state.

From an ecological point of view, the number of species of a
population must be kept above a minimum level to prevent going to
extinction due to random fluctuations of environmental conditions.
Inspired by this viewpoint, for the present MRM, we judge that the
number of the minority should not be smaller than $2$, below which
they are doomed to extinct (since two agents of the same opinion
is guaranteed to convince a neighbor), but above which they have a
chance (though very small) to survive. Thus, a reasonable
estimation is $f_{min}=f(1-2/N)$. In Fig. \ref{fig3}, we present
simulation results for three systems with different sizes,
$N=5\times10^3$, $10^4$, and $2\times10^4$, respectively. We can
see that the critical value of $q_c^\prime$ is decreasing with the
increase of the system size. For $\beta=0.9$, the simulations
yielded that $q_c^\prime=0.73$, $0.70$, and $0.66$ for
$N=5\times10^3$, $10^4$, and $2\times10^4$, respectively, which
are in good accordance with the estimations of $f(1-2/N)=0.717$,
$0.692$, and $0.665$ for the three corresponding values of $N$.

To sum up, we have investigated the effects of free will and
massive opinion of multi-agents in a network based on the majority
rule model, which is a simple yet useful statistical model for
studying the emergence of collective behaviors. Two types of
models have been considered. In model I, the agents have a free
will to either adopt the opposite opinion or remain its current
opinion, whereas in model II, besides the local majority, the
agents also have a tendency to becoming the global majority when
there are no neighboring interactions among them. We found that
the location of the order-disorder phase transition point strongly
depends on the involved dynamics, which may give rise to either
smooth or harsh conditions to achieve an ordered state. In
addition, for finite system sizes, we found that there exists a
threshold below which a full consensus can be attained if the
agents have a inclination to becoming the global majority. Our
analytical estimations are in good agreement with simulation
results. In the present work, however, we only studied the
majority rule model over fully connected networks. To model the
real world in a more accurate fashion, future research may take
into account the coevolution of the majority process and the
underlying interaction pattern, i.e., from the viewpoint of
adaptive and co-evolutionary networks \cite{Holme2006pre}.

\medskip
This work was supported by the NSFC-HK Joint Research Scheme under
the grant N-CityU 107/07.

\bibliographystyle{h-physrev3}

\begin{thebibliography}{99}
\bibitem{Note}
For any value of $\beta\geqslant 1$, it is easy to verify that the
whole system will evlove to an absorbing state where a full
consense is reached.
\end{thebibliography}

\end{document}